\documentclass[12pt]{iopart}
\usepackage{iopams}
\usepackage{graphicx}
\usepackage{cite}
\usepackage{makecell}

\newcommand{\eqref}[1]{(\ref{#1})}

\begin{document}

\title{Transition to synchrony in chiral active particles}
\author{Arkady Pikovsky$^{1,2}$}
\address{$^1$ Institute of Physics and Astronomy, University of Potsdam,
Karl-Liebknecht-Str. 24/25, 14476 Potsdam-Golm, Germany}
\address{$^2$ Control Theory Department, Institute of Information Technologies,
Mathematics and Mechanics, Lobachevsky University, 603950 Nizhny Novgorod, Russia}

\begin{abstract}
I study deterministic dynamics of chiral active particles in two dimensions.
Particles are considered as discs interacting with elastic repulsive forces.
An ensemble of particles, started from random initial conditions, 
demonstrates chaotic collisions resulting in their normal diffusion. 
This chaos is transient, as rather abruptly a synchronous 
collisionless state establishes.
The life time of chaos grows exponentially with the number of particles. 
External forcing
(periodic or chaotic) is shown to facilitate the synchronization transition. 
\end{abstract}
\maketitle

\section{Introduction}
\label{sec:intr}
Active or self-propelled particles are a subject of active current research (see introductory 
review~\cite{Bechinger_etal-16}). For microscopic particles, random interactions with an environment
are essential, and one speaks about active Brownian particles, subject to noisy forces.
For macroscopic particles, existence of random forces is not so obvious, but here one also
often introduces them to model observed fluctuations in the motions of animals and birds (cf. famous 
Vicsek model~\cite{Vicsek_etal-95}). A large class of active particles constitute \textit{chiral}
active particles, natural trajectories of which are not straight lines, but circles. The
origin of chirality can be asymmetry in the particle shape 
or in the propulsion mechanism; also external magnetic field may lead to circular
motion (see discussion and examples in~\cite{Bechinger_etal-16}). 

There are different models for interaction of active particles.
In many cases, inspired by the Vicsek model~\cite{Vicsek_etal-95},
one assumes an aligning interaction: neighboring particles ``prefer''
to align their velocities. In the context of chiral Brownian (i.e. noise-driven)
particles, such an aligning interaction can lead to an appearance of synchronized
rotating clusters (see, e.g., 
Refs.~\cite{Liao-Klapp-18,Levis-Liebchen-18,Levis-Liebchen-19,Kruk_etal-20}).
In this paper, I consider \textit{deterministic} particles \textit{without alignment
forces}. The main finding is that this system demonstrates a transition
from supertransient chaos to synchronous clusters. 

I introduce the basic model in Section~\ref{sec:mf}.  
In Section~\ref{sec:ccs} I demonstrate that while for small times,
circling colliding active particles demonstrate chaos leading to their diffusion,
at large times a transition to a synchronous state without collisions occur.
The life time of chaos grows exponentially with the number 
of particles, what allows to speak about supertransients, typical for spatio-temporal
chaos. In section~\ref{sec:dts} I demonstrate that external periodic or random forcing 
induces a transition to a collisionless state already at small times. I conclude with
discussion in Section~\ref{sec:con}.

\section{Model formulation} \label{sec:mf}
I consider active particles in two dimensions, with 
circular natural trajectories. The two components of velocity $(v=\dot x,u=\dot y)$ for one
particle obey following
equations:
\begin{equation}
\eqalign{
\dot v&=-\omega u+\mu(W^2-u^2-v^2)+m^{-1}F_x\;,\\
\dot u&=\omega v+\mu(W^2-u^2-v^2)+m^{-1}F_y\;.
}
\label{eq:bv}
\end{equation}
Here $W$ is a steady speed, which a particle attends if the other forces
$F_{x,y}$ vanish; parameter $\mu$ describes the rate of the relaxation to this steady
speed; $m$ is the particle's mass. The velocity field rotates with frequency $\omega$. In a steady state, an isolated 
particle rotates on a circle of radius $W/\omega$ with frequency $\omega$.

It is worth noting that Eq.~\eqref{eq:bv} is widely used in synchronization and coupled
oscillators studies as the Stuart-Landau model (see, e.g., \cite{Pikovsky-Rosenblum-Kurths-01}). 
In this interpretation two variables
$u,v$ describe a state of an autonomous oscillating system close to the Hopf bifurcation
point. It is also well-known
that under a periodic or random force this oscillator synchronizes~\cite{Pikovsky-Rosenblum-Kurths-01},
this effect will be explored in Section~\ref{sec:dts}.

Below I consider two types of forces acting on particles. The first
force is the interaction between the particles. I assume, following Ref.~\cite{Rex-Loewen-07}, 
a conservative repulsing
interaction governed by a truncated Lennard-Jones potential, which depends
on the distance $R$ between the particles:
\begin{equation}
V(R)=\left\{\eqalign{
\epsilon\left[\left(\frac{\sigma}{R}\right)^{12} - 2\left(\frac{\sigma}{R}\right)^6+1
\right] 
&\;\;\rm{ for }\;\; R\leq \sigma\;,\\
0&\;\;\rm{ for }\;\; R> \sigma\;.}
\right.
\label{eq:bp}
\end{equation}
This potential takes from the full Lennard-Jones potential only its repulsing part,
and the attracting part is absent. Thus, this potential mimics not-so-hard discs
with radius $\sigma/2$, which repulse each other when collide, and do not interact 
aside of collisions. I stress here that the interactions have no any alignment
action (the latter is often assumed in models of Vicseck type). If there where no
activity and chirality (i.e. $\omega=\mu=0$ in~\eqref{eq:bv}), then the model
reduces to a Hamiltonian one of elastically colliding disks. In the case of hard disks
it is believed that the dynamics is fully chaotic (the Boltzmann-Sinai hypothesis), 
although the proof~\cite{Simanyi-03} has some restrictions. For smooth potentials,
stable periodic orbits in the Hamiltonian dynamics may appear~\cite{Turaev-Rom-Kedar-03}.
Much less is known for colliding active particles, but our simulations in
Section~\ref{sec:ccs} indicate for chaos.

In addition to interaction forces described by \eqref{eq:bp}, I will consider
external forces specified in Section~\ref{sec:dts}.

Below I study numerically an ensemble of active particles with circular
orbits \eqref{eq:bv},\eqref{eq:bp} in a periodic geometry,
i.e., on a torus $L\times L$. I fix $\mu=2$, $W=\omega=1$, $\sigma=0.2$, $m=1$, and $\epsilon=0.1$
throughout the paper. The main parameters to explore is the number of the particles $N$
and the density $\rho=N L^{-2}$. 

\section{Chaotic state and spontaneous transition to synchrony}\label{sec:ccs}

\subsection{Quasistationary chaotic state}

\begin{figure}
\centering
\includegraphics[width=\columnwidth]{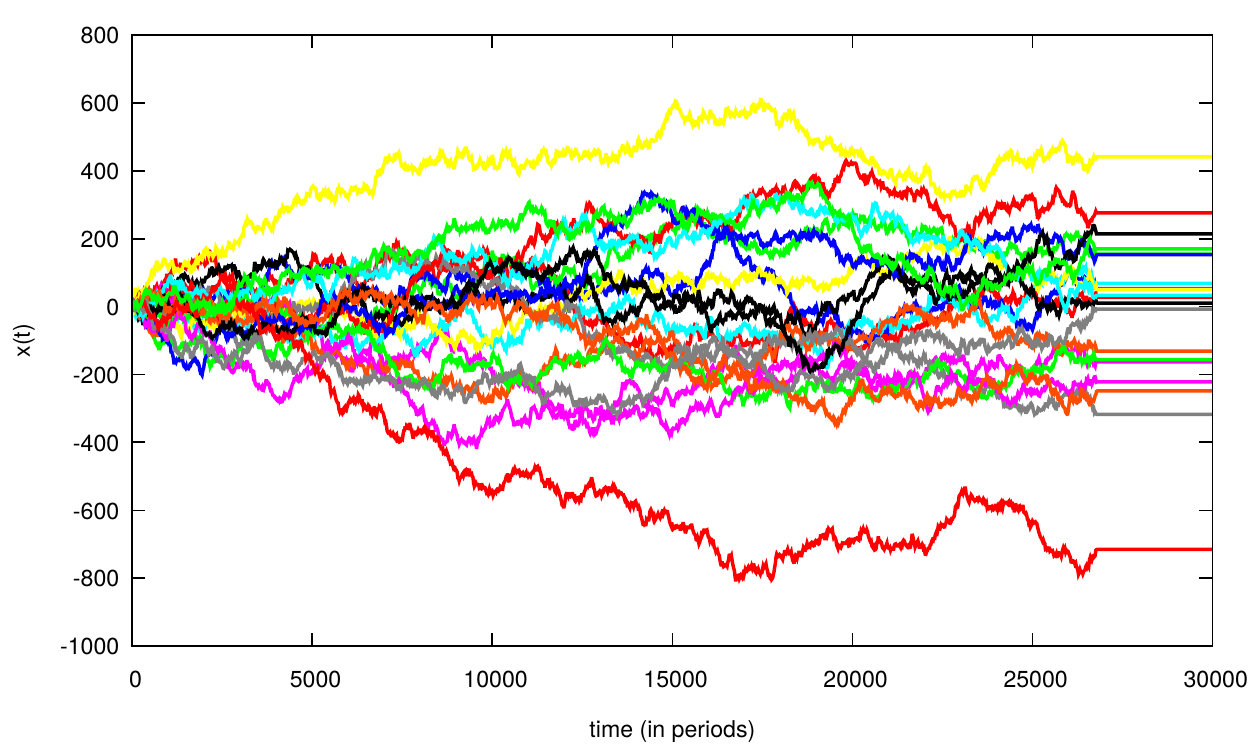}
\caption{Positions of particles vs time for $N=20$, $\rho=3$. 
Synchronization transition occurs at $t\approx 27000$. }
\label{fig:ex}
\end{figure}

\begin{figure}
\centering
\includegraphics[width=\columnwidth]{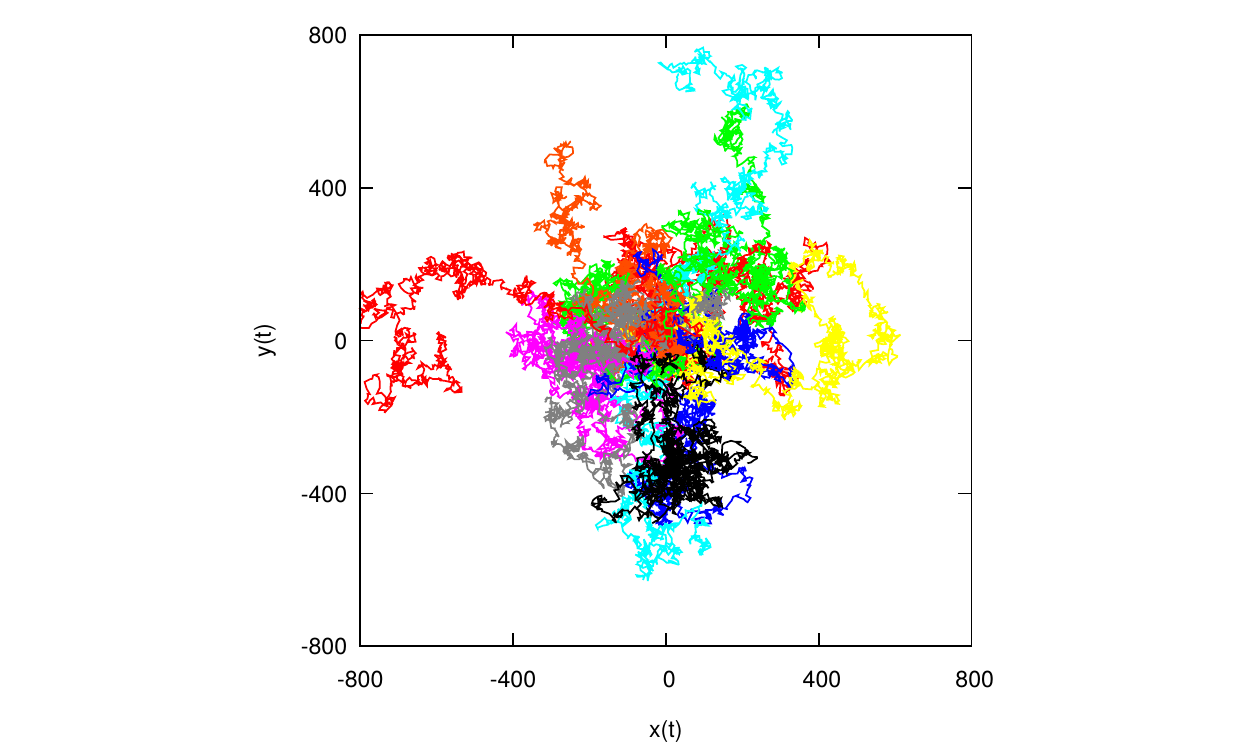}
\caption{Trajectories of particles for the same data as Fig.~\ref{fig:ex}. }
\label{fig:ex1}
\end{figure}

Because collisions of hard disk have a scattering property with
an essential degree of instability (like dispersive billiards), one can expect
chaos in
colliding active particle described by Eqs.~\eqref{eq:bv},\eqref{eq:bp}. 
Due to multiple collisions, velocity of a particle
is random, so its motion is a diffusion in two dimensions. I illustrate this
with figures \ref{fig:ex},\ref{fig:ex1}. They show trajectories of $N=20$
particles in a particular run. Up to time $T_{st}\approx 27000$ (I measure 
time in periods of rotations) motion of all particles is irregular.

\begin{figure}
\centering
\includegraphics[width=0.75\columnwidth]{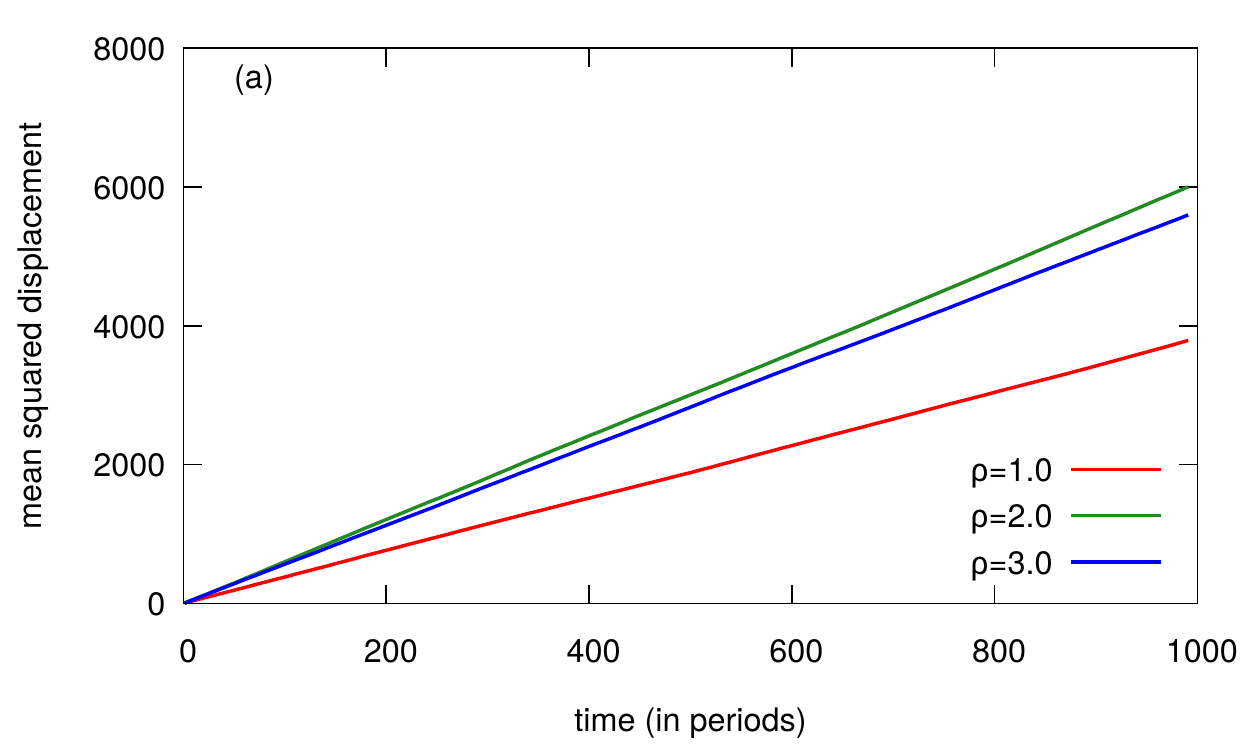}
\includegraphics[width=0.75\columnwidth]{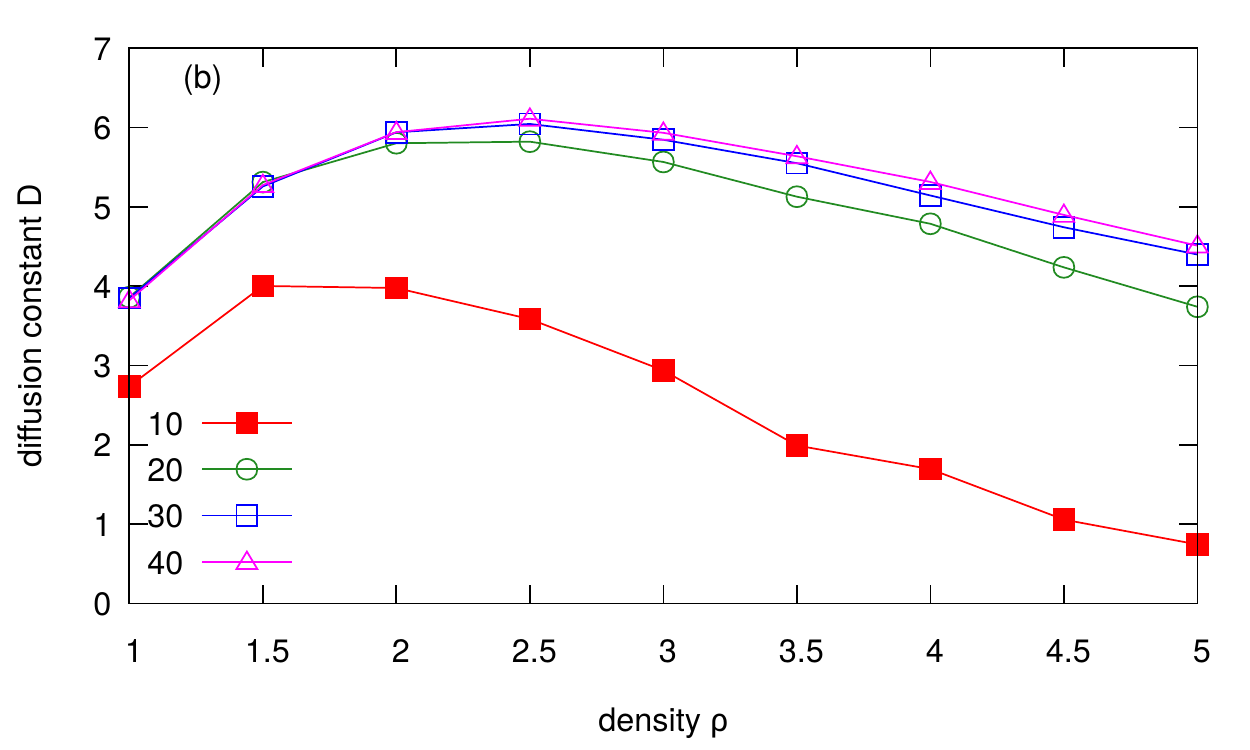}
\caption{Calculation of diffusion constant. Panel (a): mean squared displacements vs. time
for $N=20$ and different $\rho$ are nearly perfect straight lines confirming
normal diffusion of particles. Panel (b): diffusion constants for different $N$ and $\rho$.
}
\label{fig:dc1}
\end{figure}

Quantitatively, a good characteristics of irregularity is the mean diffusion constant
of the particles. It can be calculated from the mean squared displacement
after a large time interval
\[
D=\frac{\langle(x(T)-x(0))^2+(y(T)-y(0))^2 \rangle}{T}\;.
\]
Diffusion in the system is normal, as the graph of 
$\langle(x(T)-x(0))^2+(y(T)-y(0))^2 \rangle$ versus time interval $T$
(Fig.~\ref{fig:dc1}(a)) shows. Because $D$ is an intensive quantity,
one can expect that it depends on the intensive parameter - density of particles $\rho$,
but not on the number of particles $N$. However, calculations of the diffusion constant
Fig.~\ref{fig:dc1}(b) show, that for a small number of particles
a significant depletion of the diffusion constant is observed.  I attribute 
this to correlations which appear in small communities where the same 
particles collide many times. In large ensembles, a particle has again and again new neighbors,
so that correlations effectively disappear (see 
nearly coinciding values for $N=30$ and $N=40$ in Fig.~\ref{fig:dc1}(b)).


\subsection{Spontaneous transition to synchrony}
The main observation of this work is that the chaos described above is in 
fact a transient state, it evolves eventually into a configuration without collisions;
such a transition can be seen at $T_{st}\approx 27000$ in Fig.~\ref{fig:ex}.
Indeed, a set of particles \eqref{eq:bv} has an absorbing synchronous state
where all the velocities are equal: $v_1=v_2=\ldots=v_N$,  $u_1=u_2=\ldots=u_N$.
In this state the particles rotate synchronously, the distances between them remain constant,
and they do not collide. Thus this state continues forever. Such a state can exist also
in a Hamiltonian setup, but there it can occur only for specially constructed
initial conditions. If collisions in a set of Hamiltonian discs occur, they cannot disappear,
because of the reversibility of the dynamics. For active particles with a non-Hamiltonian dynamics,
there is no such a restriction.

\begin{figure}
\centering
\includegraphics[width=\columnwidth]{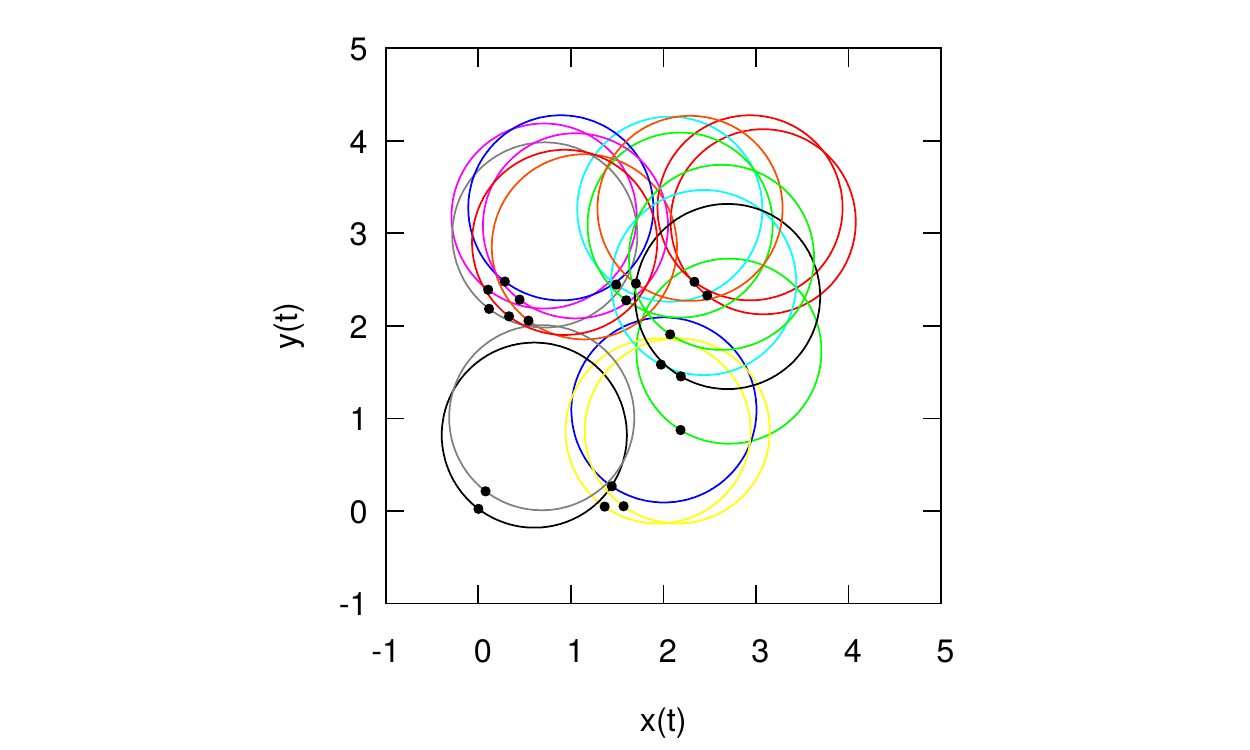}
\caption{Synchronous state for the trajectory shown in Fig.~\ref{fig:ex}. Black filled
circles: positions of the particles at a certain time; colored large circles: trajectories
of the particles.}
\label{fig:ss}
\end{figure}

I stress that the state in Fig.~\ref{fig:ex} at $T>T_{st}$ is not completely synchronous, as 
the phases of rotation of different particles do not coincide. It is sufficient to achieve
a state where collisions disappear, such a regime, which I call absorbing collisionless state, 
continues forever. I illustrate the collissionless synchronous state corresponding to
 $T>T_{st}$ of Fig.~\ref{fig:ex} in Fig.~\ref{fig:ss}.
Fig.~\ref{fig:ex} shows that the transition to synchrony is quite abrupt, so 
this is an example of type-II supertransients according to classification 
of~\cite{Lai-Tel-11}.

\begin{figure}
\centering
\includegraphics[width=\columnwidth]{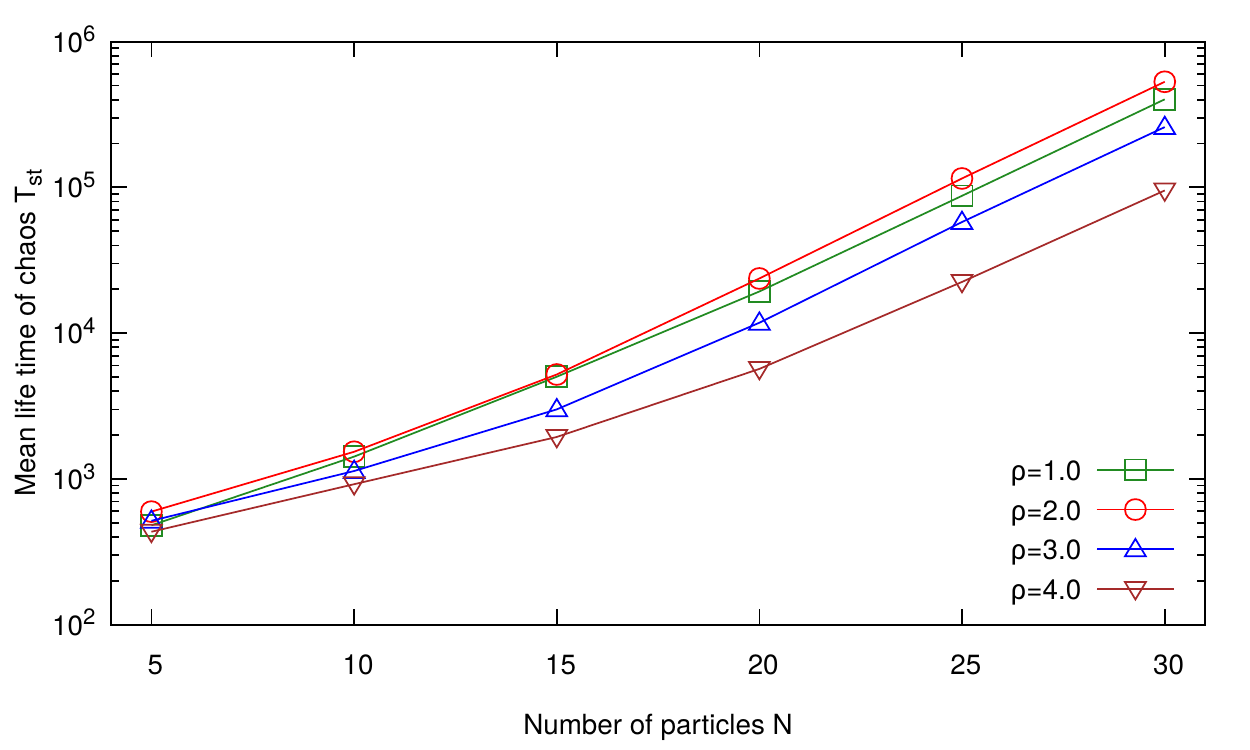}
\caption{Mean time to synchronization in dependence on the number
of particles and on density. The dependencies in the range $20\leq N\leq 30 $
are well fitted with $T_{st}\approx A(\rho)\exp[B(\rho) N]$, where $B(1)=0.304$, 
$B(2)=0.311$, $B(3)=0.309$, $B(4)=0.282$.}
\label{fig:mt}
\end{figure}

The time at which a synchronous absorbing state appears depends on the initial configuration 
of particles, it fluctuates highly. In Fig.~\ref{fig:mt} I show dependence of the mean
time to achieve a collisionless state on the number of particles and on density $\rho$.
The main feature is that the dependence on the density is rather weak, but the life time
of a chaotic state grows exponentially with the number of particles $N$. Thus,
this system belongs to a class of extended systems with chaotic 
supertransients~\cite{Crutchfield-Kaneko-88,Lilienkamp_etal-17}.
Figure \ref{fig:mt} also shows that the law of exponential growth
with the number of particles only weakly depends on the density parameter.

\section{Driven transition to synchrony}\label{sec:dts}

\begin{figure}
\centering
\includegraphics[width=\columnwidth]{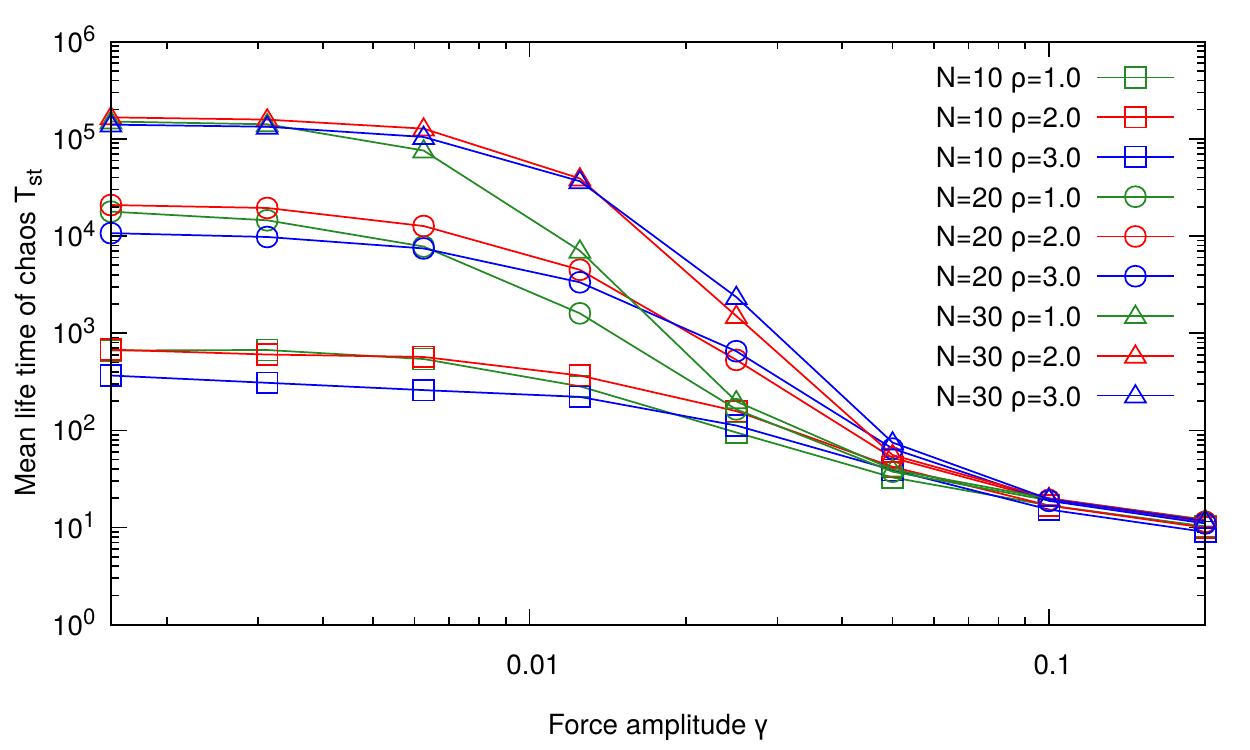}
\caption{Mean time to synchrony under periodic force vs force amplitude $\gamma$
for different $\rho$ and $N$.}
\label{fig:pf}
\end{figure}

Noninteracting particles are described by a set of effective Stuart-Landau oscillators 
\eqref{eq:bv}.
Thus, such an ensemble can be synchronized by an external periodic or random force,
as described in the theory of synchronization~\cite{Pikovsky-Rosenblum-Kurths-01}. 
One can expect that the same happens also in the presence of the collisions, although the
synchronization onset can be retarded due to them. I Illustrate the effect of a periodic force
$F_x=\gamma \cos(\omega t)$ on the ensemble in Fig.~\ref{fig:pf}. A small force has almost
no effect, the mean life time of the chaotic state is almost the same as for autonomous
particles. At large values of $\gamma$, the transition occurs within a few periods,
the mean life time is almost the same for different $N$ and $\rho$.
Noteworthy, at intermediate force amplitudes $0.01<\gamma <0.03$,
the reduction of the mean life time of chaos is mostly pronounced
for systems with low density (see curves with $\rho=1$ for $N=20$ and $N=30$).

\begin{figure}
\centering
\includegraphics[width=\columnwidth]{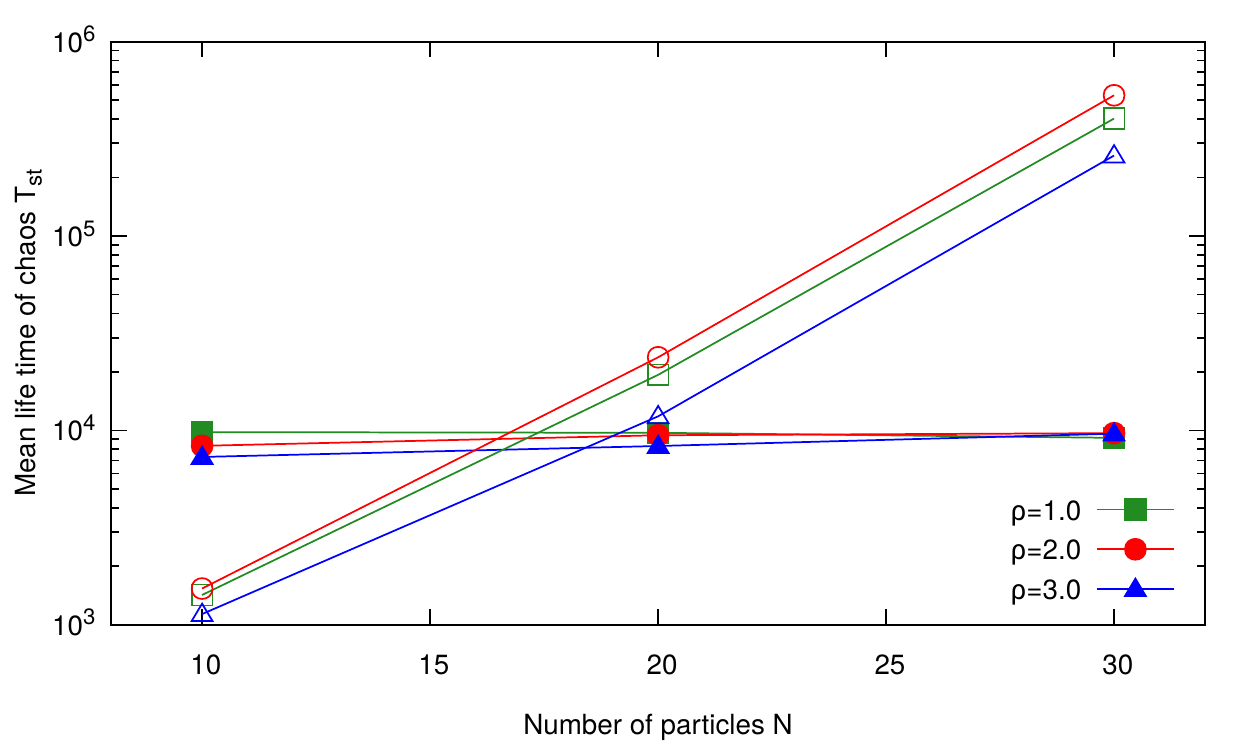}
\caption{Mean time to synchrony under random force~\eqref{eq:rf} with $\tau=0.2$
and $\sigma=0.2$ (filled markers). Also, for comparison, is shown
the unperturbed mean time (open markers).}
\label{fig:rf1}
\end{figure}

Common noise is another source of synchrony in ensembles of uncoupled
oscillators~\cite{Pikovsky-Rosenblum-Kurths-01}. Here I report on numerical
experiments where noise was in the form of a Poissonian sequence of delta-pulses:
\begin{equation}
F_x=\sum_n a_n\delta(t-t_n)\;,
\label{eq:rf}
\end{equation}
where $t_n$ are Poissonian time events appearing with rate $\tau^{-1}$, and $a_n$ are independent
amplitudes of the pulses taken from a Gaussian distribution with
standard deviation $\sigma$. Mean life times of chaos are shown in Fig.~\ref{fig:rf1}. 
Remarkably, under noisy force the mean life time of chaos is nearly a constant, it
does depend on the number of particles and on the density. Furthermore, for 
small number of particles common noise retards transition to synchrony compared to the spontaneous one.
For large number of particles the effect is opposite, here for $N=30$
noise reduces the life time of chaos by a factor larger than $20$.

\section{Conclusion}\label{sec:con}
In this paper I studied an ensemble of active particles with circular natural trajectories,
subject to elastic collisions. This setup differs from previous studies in two aspects: (i) 
the basic dynamics is purely deterministic (except for the special considered case of a common random force),
and (ii) there are no aligning interactions. I show that the autonomous system demonstrates 
supertransient chaos. Starting with random initial conditions, for a long interval of time, chaotic
dynamics is observed, leading to a normal diffusion of the particles. However, the final state
is a regime without collisions, where neighboring
particles effectively synchronize their circular rotations. I demonstrate that the life time of chaos
depends on the density of particles, and grows exponentially with the number of them. Furthermore, 
I show that the transition to synchronous rotations can be induced by an
external periodic force, common for all particles (this
effect has been previously reported in other context 
in Refs.~\cite{Chepelianskii2007,Shepelyansky2009}). Also I
demonstrate that a random common force also leads to a synchronous dynamics without collisions. 
Forced transitions
have a certain transient time which for large forces does not depend on the number of particles.

\ack
Author thanks H. Chat\'e and F. Ginelli for fruitful discussions, and the
Russian Science Foundation (studies 
of Section \ref{sec:dts}, Grant Number 19-12-00367) for support.
%
\bibliographystyle{unsrt}

\section*{Bibliography}
 \def\cprime{$'$}

\end{document}